%% file: release.tex
\documentstyle{mn}
\input mn_junk.tex

\def\hmpc{{\rm h}$^{-1}~{\rm Mpc}\,$}
\title[Public release of data]
{Public Release of N-body simulation and related data by the Virgo consortium.}
\author[C. S. Frenk et al.]
{C.~S.~Frenk,$^1$ J.~M.~Colberg,$^{2}$ H.~M.~P.~Couchman,$^3$ G.~Efstathiou,$^4$\\
\newauthor A.~E.~Evrard,$^5$ A.~Jenkins,$^1$ T.~J.~MacFarland,$^6$ B. Moore,$^1$ J.~A.~Peacock,$^7$    \\
\newauthor  F.~R.~Pearce,$^1$  P. A.~Thomas,$^8$  S.~D.~M.~White$^2$ and N.~Yoshida.$^2$ \\
\newauthor $\phantom{xxxxxxxxxxxxxxxxxxx}$(The Virgo Consortium)$^{10}$\\
\newauthor \\
$^1$ Dept Physics, South Road, Durham, DH1 3LE.\\
$^2$ Max-Planck Inst. for Astrophysics, Garching, Munich, D-85740, Germany.\\
$^3$ Department of Physics and Astronomy, McMaster University, Hamilton,
Ontario, L8S 4M1, Canada\\
$^4$ Institute of Astronomy, Madingley Road, Cambridge CB3 OHA\\
$^6$ Now at 105 Lexington Avenue, Apt. 6F,New York, NY 10016\\
$^5$ Dept Physics, University of Michigan, Ann Arbor, MI-48109-1120.\\
$^7$ Royal Observatory, Institute of Astronomy, Edinburgh, EH9 3HJ\\
$^8$ Astronomy Centre,CPES, University of Sussex, Falmer, Brighton, BN1 9QJ\\
$^{10}$ http://star-www.dur.ac.uk/$\sim$frazerp/virgo/ \\}

\pagerange{\pageref{firstpage}--\pageref{lastpage}}
\pubyear{2000}

\begin{document}

\label{firstpage}

\maketitle

\begin{abstract}

We are making available on the WWW a selection of the archived data from
N-body simulations carried out by the Virgo consortium and related
groups.  This currently includes: (i) time-slice, lightcone 
and cluster data from the two $10^9$-particle
Hubble volume simulations described by Evrard 1998; (ii) time-slice
data from simulations
of 4 different cold dark matter cosmological models with $256^3$ particles
analysed by Jenkins \etal 1998; (iii) Dark halo catalogs, merger trees and
galaxy catalogs from
the GIF project described by Kauffmann et al 1999. Basic software is supplied to
read the data. The data can be accessed from:

http://www.mpa-garching.mpg.de/Virgo/data\_download.html
\end{abstract}

\section{Introduction}

The Virgo consortium is an international collaboration whose aim is to
perform large cosmological simulations of the formation of structure. We
have been carrying out simulations on parallel supercomputers since 1996.
In this short article we provide a brief description of datasets that we
are making publicly available and we give a list of papers for refereed
journals which have already made use of the data.

Over the past few years, the Virgo consortium has carried out a large
number of high-resolution simulations of the clustering evolution of cold
dark matter (CDM). They follow structure formation in four cosmological
models: a flat model with $\Omega_0=0.3$ and cosmological constant,
$\Lambda/3H_0^2=0.7$ (\lcdm); an open model also with $\Omega_0=0.3$
(\ocdm); and two models with $\Omega=1$, one with the standard CDM power
spectrum (\scdm), and the other with the same power spectrum as the
$\Omega_0=0.3$ models (\tcdm). The number of particles in the
older simulations
is about 17 million, with the box sizes ranging from 85 to 240 \hmpc (where
${\rm h}$ denotes the Hubble constant in units of 100 km s$^{-1}$
Mpc$^{-1}$) and the particle mass from 6.86 to 22.7 $\times 10^{10}{\rm
h^{-1} M_\odot}.$
\begin{table*}
\centering
\begin{minipage}{140mm}
\caption{N-body simulation parameters. The Hubble volume data, with
-hub suffix,  includes
lightcone outputs as well as time-slices (see Evrard 1998).  The 
$N=256^3$ simulations were first analysed by Jenkins \etal 
(1998).  The simulations with the -gif suffix were designed
to allow galaxy formation to be followed directly and were first
fully presented  (including the available halo and galaxy catalogs and 
merger trees) by Kauffmann \etal  (1999).  }
\begin{tabular}{@{}lrrlrrrrr@{}}
Run  &  $\Omega_0$ & $\Lambda_0$ & $\Gamma$ & $\sigma_8$ & 
$N_{\rm part}\phantom{aaa}$ 
     & $L_{\rm box}/h^{-1}{\rm Mpc}$ & $m_{\rm particle}/h^{-1}\msun$ & 
$r_{\rm soft}/h^{-1}{\rm kpc}$\\
\tcdm-hub &   1.0  & 0.0 & 0.21 & 0.60 & $1\;000\,000\,000$ & $2000.0\phantom{bbbb}$ 
& $2.22\times10^{12}\phantom{ccc}$& $100\phantom{aaaaaa}$\\
\lcdm-hub & 0.3  & 0.7 & 0.21& 0.90 & $1\,000\,000\,000$ & 
$3000.0\phantom{bbbb}$ & $2.25\times10^{12}\phantom{ccc}$& $100\phantom{aaaaaa}$\\
\scdm-virgo &   1.0  & 0.0 & 0.5 & 0.60 & $16\,777\,216$ & $239.5\phantom{bbbb}$ 
& $2.27\times10^{11}\phantom{ccc}$& $30\phantom{aaaaaa}$\\
\tcdm-virgo &   1.0  & 0.0 & 0.21 & 0.60 & $16\,777\,216$ & $239.5\phantom{bbbb}$ 
& $2.27\times10^{11}\phantom{ccc}$& $30\phantom{aaaaaa}$\\
\lcdm-virgo &   0.3  & 0.7 & 0.21 & 0.90 & $16\,777\,216$ & $239.5\phantom{bbbb}$ 
& $6.86\times10^{10}\phantom{ccc}$& $30\phantom{aaaaaa}$\\
\ocdm-virgo &   0.3  & 0.0 & 0.21 & 0.85 & $16\,777\,216$ & $239.5\phantom{bbbb}$ 
& $6.86\times10^{10}\phantom{ccc}$& $30\phantom{aaaaaa}$\\
\scdm-gif &   1.0  & 0.0 & 0.5 & 0.60 & $16\,777\,216$ & $84.5\phantom{bbbb}$ 
& $1.00\times10^{10}\phantom{ccc}$ & $30\phantom{aaaaaa}$\\
\tcdm-gif &   1.0  & 0.0 & 0.21 & 0.60 & $16\,777\,216$ & $84.5\phantom{bbbb}$ 
& $1.00\times10^{10}\phantom{ccc}$ & $30\phantom{aaaaaa}$\\
\lcdm-gif & 0.3  & 0.7 & 0.21& 0.90 & $16\,777\,216$ & 
$141.3\phantom{bbbb}$ & $1.40\times10^{10}\phantom{ccc}$& $30\phantom{aaaaaa}$\\
\ocdm-gif & 0.3  & 0.0 & 0.21& 0.85 & $16\,777\,216$ & 
$141.3\phantom{bbbb}$ & $1.40\times10^{10}\phantom{ccc}$& $30\phantom{aaaaaa}$\\
\end{tabular}
\end{minipage}
\end{table*}

A variety of studies have been carried out using these data.  These include
investigations of the clustering statistics of the dark matter
(Jenkins \etal 1998, Juskiewicz, Springel \& Durrer 1999); detailed 
analyses of the internal structure (Thomas
\etal 1998), velocity distributions (Colberg \etal 2000) and the
large-scale environment of galaxy cluster halos (Colberg \etal 1999); a
novel description of the topology of the dark matter distribution and a
comparison with recent galaxy surveys (Springel \etal 1998, Canavezes
\etal  1998, Canavezes \& Sharpe 1998); analyses of the strong 
gravitational lensing properties of
clusters (Bartelmann \etal 1998); ray-tracing simulations and
statistical analysis of the shear fields produced by weak lensing
(Reblinsky \& Bartelmann 1999, Reblinsky et al 1999, Jain, Seljak \&
White 2000, Jain \& van Waerbecke 2000, Munshi \& Jain 2000a,b);
studies of the Sunyaev-Zel'dovich signatures imprinted on the CMB
(Diaferio, Sunyaev \& Nusser 2000); tests of analytic models both for the
abundance and clustering of dark halos (Sheth \& Tormen 1999;
Sheth, Mo \& Tormen 2000) and for the bias of the galaxy population
(Peacock \& Smith 2000).

For the higher resolution simulations the relation between the galaxy
and mass distributions has been studied both by inserting
``semi-analytic'' galaxy populations into the simulated dark halos at
various times (Benson \etal 2000a,b) and by using similar ``recipes''
for baryonic processes to follow the formation of galaxies directly in
the simulations (the GIF project: Kauffmann \etal 1999ab, Diaferio
\etal 1999, 2000; Somerville \etal 2000). Galaxy populations
constructed with the first technique have also been applied to the
study of cosmic reionization (Benson \etal 2000c) and in making
predictions of clustering and higher order correlations which are
compared to measurements from the PSCz survey (Szapudi \etal
2000). The GIF galaxy, mass and halo catalogs have been used by others
for studies of the statistics of the galaxy distribution (Seljak 2000,
Scoccimarro \etal 2000), of lensing effects due to individual galaxies
(Guzik \& Seljak 2000), of infall patterns onto galaxy clusters
(Diaferio 1999) and for comparison with the topology of the local
galaxy distribution (Schmalzing \& Diaferio 2000).

Several years ago we completed the two largest N-body simulations ever
attempted using a specially designed code (MacFarland \etal  1999). These
``Hubble volume" simulations follow a billion particles each -- almost one
order of magnitude more than the largest previous simulations -- in the
\tcdm\ and \lcdm\ cosmologies. 
The simulated regions have linear, comoving dimensions of 2000-3000 \hmpc
so that, for $\Omega=1$, the diagonal of the simulation cube extends to a
redshift $z=4.6$, nearly the size of the ``observed'' universe. The total
mass in the $\Omega=1$ volume is $2.2
\times 10^{21} {\rm h}^{-1} {\rm M_\odot}$, and so a cluster as rich as
Coma has about 500 particles within an Abell radius. The comoving
gravitational softening, $0.1 {\rm h}^{-1} {\rm Mpc}$, is sufficiently
small to provide reliable cluster velocity dispersions and gross
assembly histories. Particles masses and softenings are similar in the
\lcdm\ simulation. Such large volumes require the application
of new analysis techniques that take into account the substantial light
travel time across the region. In addition to storing a number of standard 
``snapshots'' of the mass distribution, we therefore created surveys along the
past light cones of hypothetical observers.

The Hubble volume simulations contain several tens of thousands of rich
clusters and are so large that they can be used to investigate the
behaviour of clustering diagnostics, including high order statistics, with
unprecedented accuracy. The first papers analyzing Hubble volume data
(Szapudi \etal 2000, Colombi \etal 2000) present the probability
distribution of fundamental statistics such as the distribution of
counts-in-cells, together with a full analysis of sampling
uncertainties for estimates constructed from surveys of realistic
(finite) size.  Analysis of the two-point correlations of
clusters is presented in Colberg \etal  (2000). Data from the Hubble 
volume simulations, together with a wide range of other Virgo data, has
been used to study the mass function of dark matter halos over
more than four orders of magnitude in mass (Jenkins \etal  2000).
Mock-galaxy catalogs constructed from the Hubble Volume simulations 
are used to assess the uncertainties in observational power spectrum
estimates by Hoyle \etal  (1999).

The numerical data from all the original $N=256^3$ Virgo simulations 
are now publically available, as are a variety of data products from the
Hubble Volume simulations (``snap-shots'', light-cone output and
cluster catalogs)and from the galaxy formation modelling
of the GIF project (halo catalogs, merger trees and galaxy catalogs). 
They may be found at: 
http://www.mpa-garching.mpg.de/

\noindent Virgo/data\_download.html  

\noindent{We anticipate making further data available from
this site as our projects progress. Smaller data volumes can be
down-loaded directly over the net. Many of our datasets are too large,
however, for this to be practicable. Such data can be obtained on tape
after payment of a nominal fee; we impose this to avoid frivolous
requests imposing unnecessary work on our archive manager. Images of
many of these simulations can be down-loaded from the related websites:}

http://www.mpa-garching.mpg.de/NumCos

http://star-www.dur.ac.uk/$\sim$frazerp/virgo/

http://www.physics.lsa.umich.edu/hubble-volume/

We would appreciate an appropriate credit if these data are used in
published work.  This should include the relevant reference as
detailed below, as well as a mention in the acknowledgements that the
data were obtained from the Virgo consortium.  The $N=256^3$
simulations in Table 1 were first presented in Jenkins et al
(1998). The original motivation for the set-up of the -gif simulations
and the additional analysis involved in constructing halo catalogs,
merger trees and galaxy catalogs are described in Kauffmann \etal
(1999a). The Hubble Volume simulations are outlined in Evrard (1998)
and a detailed description is in preparation (Evrard etal 2000). We
would like to keep an up-to-date record of all published papers using
our data and so we would appreciate your help in notifying us by email
(to C.S.Frenk@durham.ac.uk) of new papers making use of Virgo data.

\section*{Acknowledgements}

CSF acknowledges a PPARC Senior Research Fellowship and a Leverhulme
Research Fellowship. PAT is a PPARC Lecturer Fellow.  AEE's
participation was supported in France by the CIES and CNRS and in the
US by NSF AST98-03199 and by the NASA Astrophysics Theory Program.
The simulation data were generated using supercomputers at the
Computer Centre of the Max-Planck Society (the Rechenzentrum,
Garching) and at the Edinburgh Parallel Computing Centre. We
particularly wish to thank Jacob Pichlmeier (SGI/Cray, RZG) for his
support. We thank the Max-Planck-Institute for Astrophysics for
hosting our down-load site and, in particular, Norbert Gruener for
acting as site administrator.

\label{lastpage}

\clearpage
 
\label{lastpage}
\end{document}

%% file: mn_junk.tex
\newif\ifAMStwofonts

\newcommand{\figstart}[1]  {\begin{figure} \psfig{#1}}
\newcommand{\lfigstart}[1] {\begin{figure*} \psfig{#1}}
\newcommand{\figend} {\end{figure}}
\newcommand{\lfigend} {\end{figure*}}

\ifoldfss
  \ifCUPmtlplainloaded \else
    \NewTextAlphabet{textbfit} {cmbxti10} {}
    \NewTextAlphabet{textbfss} {cmssbx10} {}
    \NewMathAlphabet{mathbfit} {cmbxti10} {} 
    \NewMathAlphabet{mathbfss} {cmssbx10} {} 
  \fi
  \ifAMStwofonts
    \ifCUPmtlplainloaded \else
      \NewSymbolFont{upmath} {eurm10}
      \NewSymbolFont{AMSa} {msam10}
      \NewMathSymbol{\upi}     {0}{upmath}{19}
      \NewMathSymbol{\umu}     {0}{upmath}{16}
      \NewMathSymbol{\upartial}{0}{upmath}{40}
      \NewMathSymbol{\leqslant}{3}{AMSa}{36}
      \NewMathSymbol{\geqslant}{3}{AMSa}{3E}

    \fi
  \fi
\fi 

\ifnfssone
  \newmathalphabet{\mathit}
  \addtoversion{normal}{\mathit}{cmr}{m}{it}
  \addtoversion{bold}{\mathit}{cmr}{bx}{it}
  \newmathalphabet{\mathbfit} 
  \addtoversion{normal}{\mathbfit}{cmr}{bx}{it}
  \addtoversion{bold}{\mathbfit}{cmr}{bx}{it}
  \newmathalphabet{\mathbfss} 
  \addtoversion{normal}{\mathbfss}{cmss}{bx}{n}
  \addtoversion{bold}{\mathbfss}{cmss}{bx}{n}
  \ifAMStwofonts
    \ifCUPmtlplainloaded \else
      \UseAMStwoboldmath
      \makeatletter
      \new@mathgroup\upmath@group
      \define@mathgroup\mv@normal\upmath@group{eur}{m}{n}
      \define@mathgroup\mv@bold\upmath@group{eur}{b}{n}
      \edef\UPM{\hexnumber\upmath@group}
      \new@mathgroup\amsa@group
      \define@mathgroup\mv@normal\amsa@group{msa}{m}{n}
      \define@mathgroup\mv@bold\amsa@group{msa}{m}{n}
      \edef\AMSa{\hexnumber\amsa@group}
      \makeatother
      \mathchardef\upi="0\UPM19
      \mathchardef\umu="0\UPM16
      \mathchardef\upartial="0\UPM40
      \mathchardef\leqslant="3\AMSa36
      \mathchardef\geqslant="3\AMSa3E
    \fi
  \fi
\fi 

\ifnfsstwo
  \DeclareMathAlphabet{\mathbfit}{OT1}{cmr}{bx}{it}
  \SetMathAlphabet\mathbfit{bold}{OT1}{cmr}{bx}{it}
  \DeclareMathAlphabet{\mathbfss}{OT1}{cmss}{bx}{n}
  \SetMathAlphabet\mathbfss{bold}{OT1}{cmss}{bx}{n}
  \ifAMStwofonts
    \ifCUPmtlplainloaded \else
      \DeclareSymbolFont{UPM}{U}{eur}{m}{n}
      \SetSymbolFont{UPM}{bold}{U}{eur}{b}{n}
      \DeclareSymbolFont{AMSa}{U}{msa}{m}{n}
      \DeclareMathSymbol{\upi}{0}{UPM}{"19}
      \DeclareMathSymbol{\umu}{0}{UPM}{"16}
      \DeclareMathSymbol{\upartial}{0}{UPM}{"40}
      \DeclareMathSymbol{\leqslant}{3}{AMSa}{"36}
      \DeclareMathSymbol{\geqslant}{3}{AMSa}{"3E}
    \fi
  \fi
\fi 

\ifCUPmtlplainloaded \else
  \ifAMStwofonts \else 
    \def\upi{\pi}
    \def\umu{\mu}
    \def\upartial{\partial}
  \fi
\fi

\def\hmpc{h^{-1}{\rm Mpc}}

\def\msun{M_\odot}

\def\lcdm{$\Lambda$CDM}
\def\tcdm{$\tau$CDM}
\def\scdm{SCDM}
\def\ocdm{OCDM}
\def\etal{{et al\ }}